\documentstyle[epsf,prl,aps]{revtex}
\begin{document}
\twocolumn[\hsize\textwidth\columnwidth\hsize\csname@twocolumnfalse\endcsname
\preprint{\vbox{\hbox{November 1997} \hbox{IFP-746-UNC}  }}
\title{\bf $S$ and $T$ for Leptoquarks and Bileptons}
\author{\bf Paul H. Frampton and Masayasu Harada}
\address{Department of Physics and Astronomy, University of North Carolina,\\
Chapel Hill, NC 27599-3255.}
\maketitle

\begin{abstract}
We calculate contributions to
the oblique parameters $S$ and $T$ 
for leptoquarks and bileptons, and
find phenomenological constraints on their
allowed masses. 
Leptoquarks suggested by the neutral and charged 
current anomalies at HERA can give improved agreement
with both $S$ and $T$. 
If bileptons are the only new states,
the singly-charged one must be heavier than
the directly-established lower limit.
We also consider $SU(15)$
grand unification to find 
when it can be consistent
with precision electroweak measurements.  
\end{abstract}
\vskip1pc]

Model building for particle theory at and below the TeV
energy scale is not rendered redundant by the standard
model but is certainly very constrained by it. There
can be little that we do not already know much below the $W$ mass.
In the mass region up to the TeV regime
one expects the Higgs boson and perhaps supersymmetric partners. 
But otherwise one 
must tread softly to avoid upsetting the delicate agreement
between the standard model and precision experimental data.

In the present paper we take a fresh look at
how certain additional bosons can co-exist
with one another and with the successful
standard model. Of special interest are particles
which can contribute negatively to the very useful
parameters $S$ and $T$~\cite{st}
 which measure compatibility
with precision electroweak data. This is because
these parameters are generally positivite 
especially for the majority of fermionic
particles which one might add.
We shall focus on bileptonic gauge bosons and scalar 
leptoquarks. 

The H1~\cite{h1} and ZEUS~\cite{zeus} collaborations reported a
possible excess of neutral current (NC) events in $e^+p$ collisions.
This excess can be explained by the existence of the leptoquark of
mass around $200$\,GeV\cite{lq}.
In addition to the NC channel,
an excess in the charged current (CC) was also
reported\cite{Krakauer}.

Recently, D0~\cite{d0} and CDF~\cite{cdf} gave the 95\% C.L. lower
limit on the masses of 225\,GeV and 213\,GeV with assuming unit
branching fraction into a first-generation charged lepton plus jet.
For compatibility with the HERA anomaly, therefore, the scalar leptoquark
must have a significant branching ratio to other decay channels.

In Ref.~\cite{BKM} a mixing between two scalar leptoquark doublets
carrying different weak hypercharges ($Y=7/6$ and $1/6$)
was introduced to simultaneously
explain the NC and CC anomaly at HERA.
Since the lightest leptoquark couples to both $e+j$ and $\nu+j$, the
CDF/D0 limits are weakened.
They also studied the contributions to $\rho$ parameter~\cite{veltman},
and showed that $\Delta\rho$ from leptoquark could be negative.
Since a relatively large mixing is needed, another electroweak
precision parameter $S$~\cite{st} should be also studied.  [The
parameter $T$ is equivalent to $\Delta\rho$.]

Let us write these doublets as
$\Phi_{7/6}$ ($Y=7/6$) and $\Phi_{1/6}$ ($Y=1/6$), both of which
belong to a {\bf $3$} representation of SU(3)$_{\rm C}$.
The electric charges are 
$\Phi_{7/6}(Q = 5/3, 2/3)$ and 
$\Phi_{1/6}(Q = 2/3, - 1/3)$. 
The SU(3)$_{\rm C}\times$SU(2)$_{\rm L}\times$U(1)$_{\rm Y}$ invariant
mass terms are given by
\begin{equation}
{\cal L}_M = - M^2 \Phi^{\dag}_{7/6} \Phi_{7/6}
- {M'}^2 \Phi^{\dag}_{1/6} \Phi_{1/6} \ .
\label{inv mass}
\end{equation}
The interactions to the standard Higgs field $H$ are given by
\begin{eqnarray}
{\cal L}_H &=&
- \lambda_1 \left\vert H^{\dag} \Phi_{7/6} \right\vert^2
- \lambda_2 \left\vert H^{\dag} \Phi_{1/6} \right\vert^2
\nonumber\\
&&
- \tilde{\lambda}_1 
  \left\vert \tilde{H}^{\dag} \Phi_{7/6} \right\vert^2
- \tilde{\lambda}_2 
  \left\vert \tilde{H}^{\dag} \Phi_{1/6} \right\vert^2
\nonumber\\
&&
- \lambda_3 \left[
  (\Phi_{7/6})^{\dag} H \tilde{H}^{\dag} \Phi_{1/6}
  + {\rm h.c.}
\right]
\ , 
\label{Higgs}
\end{eqnarray}
where $\tilde{H}=(i\tau_2 H)^T$.\footnote{
There are other terms like 
$\vert H^{\dag} H\vert \Phi_{7/6}^{\dag} \Phi_{7/6}$.  But their
contributions are absorbed into the redefinitions of $M$ and $M'$ in
Eq.~(\ref{inv mass}).
}
After electroweak symmetry breaking by
the vacuum expectation value (VEV) of $H$,
the $\lambda_1$ and $\tilde{\lambda}_1$ terms give mass splittings
between $Q=5/3$ and $Q=2/3$ components
$(\Phi_{7/6}^{5/3},\Phi_{7/6}^{2/3})$ of $\Phi_{7/6}$, and $\lambda_2$
and $\tilde{\lambda}_2$ terms make mass difference between
$\Phi_{1/6}^{2/3}$ and $\Phi_{1/6}^{-1/3}$.
On the other hand, the $\lambda_3$ term gives mixing between two
$Q=2/3$ leptoquarks of $\Phi_{7/6}$ and $\Phi_{1/6}$.
Let $\alpha$ denote this mixing angle:
\begin{equation}
\left(\begin{array}{c}
  \Phi^{2/3}_l \\ \Phi^{2/3}_h
\end{array}\right)
=
\left(\begin{array}{cc}
  \cos\alpha & -\sin\alpha \\
  \sin\alpha & \cos\alpha  
\end{array}\right)
\left(\begin{array}{c}
  \Phi^{2/3}_{7/6} \\ \Phi^{2/3}_{1/6}
\end{array}\right)
\ ,
\end{equation}
where $\Phi^{2/3}_l$ and $\Phi^{2/3}_h$ denote mass eigenstates.
[We use the convention where the mass of $\Phi^{2/3}_l$ is lighter
than that of $\Phi^{2/3}_h$.]
We do not write explicit form of mases
$(m_{5/3},m_{-1/3},m_{2/3h},m_{2/3l})$ and the mixing angle $\alpha$
in terms of original parameters in Eqs.~(\ref{inv mass}) and
(\ref{Higgs}), since all of them are independent parameters.  [Note
that originally there are seven parameters.]  However, it should be
noticed that for the consistency of the model any differences among
all the masses are less than a few 100\,GeV~\cite{BKM}.

\begin{figure}[h]
\begin{center}
\epsfxsize=2.8in
\epsfysize=2.3in
\ \epsfbox{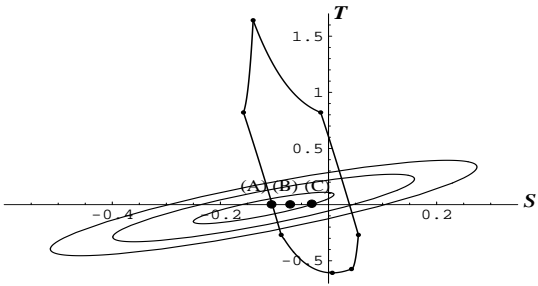}
\end{center}
\caption[]{
The possible region of values of $S$ and $T$ for 250\,GeV $\le$
$m_{5/3}$, $m_{-1/3}$, $m_{2/3h}$ $\le$ 350\,GeV with
$m_{2/3l}=200$\,GeV and $\alpha=\pi/4$. 
Three points indicated by (A), (B) and (C) correspond to the parameter
choices $(m_{2/3h},m_{-1/3},m_{5/3})=(325,250,350)$, $(350,350,350)$
and $(250,250,250)$, respectively.
}\label{fig:1}
\end{figure}

The contributions from a single SU(2)$_{\rm L}$ doublet of leptoquarks
to $S$ and $T$ are studied in Ref.~\cite{Keith-Ma}.  It is shown that
the contribution to $S$ can be negative while that to $T$ is positive
semidefinite.  However, in the present case, the situation is
different due to the existence of relatively large mixing between two
doublets.  The contributions to $S$ and $T$ from two doublets with
mixing are studied in~\cite{Lavoura-Li}.
In the present case the leptoquarks give
negative contributions to both $S$ and $T$ for a reasonable range of
parameters.  
As an example, we show in Fig.~\ref{fig:1} the possible
region of values of $S$ and $T$ for 250\,GeV $\le$ $m_{5/3}$,
$m_{-1/3}$, $m_{2/3h}$ $\le$ 350\,GeV with $m_{2/3l}=200$\,GeV and
$\alpha=\pi/4$ together with the experimental fit~\cite{Hagiwara}
for Higgs mass $M_H=300$\,GeV.
As can be read from this figure the values of $S$ and $T$ from
leptoquarks for reasonable parameter choices
agree with experiment better than the standard model.
The contours in Fig 1 (and 2) are for 39\%, 90\%, 99\%
confidence levels respectively.

In certain extensions of the standard model, there occur
bileptonic gauge 
bosons~\cite{Frampton-Lee,Frampton-Kephart,331} 
which typically occur in
$SU(2)$ doublets $(Y^{--}, Y^-)$ and the conjugates
$(Y^{++}, Y^+)$.
The experimental data currently constrain the masses
of bileptons differently for the singly-charged and doubly-charged
varieties. 
(For a recent review, see, e.g., Ref.~\cite{Cuypers}.)

The best lower limit on the singly-charged
bilepton is presently given by the precision data on
the decay of polarized muons. If the normal decay
$\mu^- \rightarrow e^- + \bar{\nu}_e + \nu_{\mu}$
is contaminated by the bilepton-mediated
$\mu^- \rightarrow e^- + \nu_e + \bar{\nu}_{\mu}$
Fierz rearrangement means that the latter contributes
proportionally to $V + A$ rather than $V - A$.
The limit on the Michel parameter $\xi$ in the 
coupling $V - \xi A$ of $\xi > 0.997$ found
at the PSI experiment\cite{PSI} gives the limit
$M(Y^-) > 230$GeV~\cite{mudecay}.

For the doubly-charged bilepton $Y^{--}$ a tighter
lower bound has been found recently from
muonium - antimuonium conversion limits, also at PSI~\cite{muonium}.
The data require that $M(Y^{--}) > 360$GeV.

In the models which predict bileptonic gauge bosons
SU(2)$_{\rm L}\times$U(1)$_{\rm Y}$ is part of a gauged 
SU(3)$_l$, and the bileptonic gauge bosons become massive
when the SU(3)$_l$ is broken.
There are generally two types of models for this kind of bileptons:
(1) bilepton gauge bosons couple to only leptons 
as in SU(15)\cite{Frampton-Lee,Frampton-Kephart};
(2) bilepton gauge bosons couple to quarks as well as leptons 
as in 3-3-1 model\cite{331}.

In the case (1) the usual SU(2)$_{\rm L}$ gauge bosons are certain
linear combinations of gauge bosons of the unbroken SU(2) subgroup of
SU(3)$_l$ and other gauge bosons coupling to quarks.
The generators of SU(2)$_{\rm L}$ and U(1)$_{\rm Y}$,
$I^a$ and $Y$, are embedded as (a = 1,2,3)
\begin{equation}
I^a = T_l^a +\cdots \ ,  \qquad
Y = -\sqrt{3} T_l^8 + \cdots \ ,
\label{generator relation}
\end{equation}
where $T_l^a$ denote the generators of SU(3)$_l$,
and dots stand for other contributions.
The relations of the gauge bosons $W_\mu^a$ and $B_\mu$ of the standard
SU(2)$_{\rm L}\times$U(1)$_{\rm Y}$ to the $A^a_l$ of SU(3)$_l$
are given by
\begin{eqnarray}
g_l A_\mu^a &=& g W_\mu^a + \cdots \ , \qquad (a=1,2,3) \nonumber\\
g_l A_\mu^8 &=& - \sqrt{3} g' B_\mu + \cdots \ ,
\label{gauge field relation}
\end{eqnarray}
where $g$, $g'$ and $g_l$ denote the corresponding gauge coupling
constants.
The bileptonic gauge bosons are expressed as
\begin{equation}
Y^{\pm\pm}_\mu = \frac{1}{\sqrt{2}}
\left( A_\mu^4 \mp A_\mu^5 \right) \ , \qquad
Y^{\pm}_\mu = \frac{1}{\sqrt{2}}
\left( A_\mu^6 \mp A_\mu^7 \right) \ .
\label{def:bilepton}
\end{equation}

In the models of type (2)
the unbroken SU(2) subgroup of 
SU(3)$_l$ is nothing but the electroweak SU(2)$_{\rm L}$.
Then the first relations of Eqs.~(\ref{generator relation})
and (\ref{gauge field relation}) become
\begin{equation}
I^a = T_l^a \ , \qquad 
g_l A_\mu^a = g W_\mu^a \ , \qquad (a=1,2,3) \ ,
\end{equation}
with $g_l=g$.
The dots parts in the second relations of 
Eqs.~(\ref{generator relation})
and (\ref{gauge field relation}) are modified.
The definitions of bileptonic gauge fields in Eq.~(\ref{def:bilepton})
remain intact.

In both types of models the bileptonic gauge bosons of SU(3)$_l$ makes
SU(2)$_{\rm L}$ doublet with hypercharge 3/2 and its conjugate.
It is convenient to use SU(2)$_{\rm L}$ doublet notation 
$Y_\mu\equiv(Y^{++}_\mu,Y^{+}_\mu)$.
The effective Lagrangian for bileptonic gauge bosons at the scale
below SU(3)$_l$ breaking scale can be written as
\begin{eqnarray}
{\cal L}_{0} &=& - \frac{1}{2} (Y_{\mu\nu})^{\dag} Y^{\mu\nu}
+ (D_\mu \Phi - i M Y_\mu)^{\dag} (D^\mu \Phi - i M Y^{\mu})
\nonumber\\
&& 
- i  g Y_\mu^{\dag} F^{\mu\nu}(W) Y^\mu
+ i \frac{3}{2}  g' Y_\mu^{\dag} F^{\mu\nu}(B) Y^\mu
\ ,
\label{LY0}
\end{eqnarray}
where $\Phi$ are the would-be Nambu-Goldstone bosons eaten by bileptonic
gauge bosons: $\Phi = (\Phi_{++},\Phi_{+})$.
The $Y_{\mu\nu}$ and $D_\mu\Phi$ are given by
\begin{eqnarray}
Y_{\mu\nu} &=&
D_\mu Y_\nu - D_\nu Y_\mu \ ,
\nonumber\\
D_\mu \Phi &=&
\left[
  \partial_\mu - i g W_\mu + i \frac{3}{2} g' B_\mu
\right]
\Phi
\ .
\end{eqnarray}

In the simplest case the SU(2)$_{\rm L}$ doublet Higgs field
is introduced as a part of SU(3)$_l$ triplet (or anti-triplet).
The other component field generally carries lepton number two,
and SU(2)$_{\rm L}$ singlet with hypercharge one or two:
\begin{equation}
\phi = \left(
\begin{array}{c}
H_1 \\ \phi_-
\end{array}
\right)
\ , \qquad
\phi' = \left(
\begin{array}{c}
H_2 \\ \phi_{--}
\end{array}
\right)
\ ,
\end{equation}
where $H_1$ ($H_2$) is a SU(2)$_{\rm L}$ doublet field with 
hypercharge $1/2$ ($-1/2$), and $\phi_-$ ($\phi_{--}$) is a 
SU(2)$_{\rm L}$ singlet field with hypercharge $-1$ ($-2$).

Both the VEVs of these Higgs fields $H_1$ and $H_2$
give masses to $W$ and $Z$ bosons, and the standard electroweak
SU(2)$_{\rm L}\times$U(1)$_{\rm Y}$ is broken.
The VEV of $H_1$ gives a mass correction to $Y^-$, while
the VEV of $H_2$ gives a mass correction to $Y^{--}$.
If only one Higgs doublet had VEV, the mass
difference of bileptons would be related to the mass of $W$ boson.
But in realistic models several Higgs fileds are needed to have VEVs.
In such a case both the masses of bileptons and $W$ boson are
independent with each other.  
In the following we regard the bilepton masses as independent
quantities.
Moreover, the actual would-be Nambu-Goldstone bosons eaten by
bileptonic gauge bosons are certain linear combinations of $\Phi$ in
Eq.~(\ref{LY0}) with $\phi_-$ or $\phi_{--}$.
We assume that the contributions to $S$ and $T$ due to these mixings
are small compared with the bilepton contributions.
Thus we use the following effective Lagrangian for the
kinetic term of would-be Nambu-Goldstone bosons and bilepton masses
after SU(2)$_{\rm L}\times$U(1)$_{\rm Y}$ is broken:
\begin{equation}
{\cal L}_{\rm NG} =
(D_\mu \Phi - i \hat{M} Y_\mu)^{\dag} (D^\mu \Phi - i \hat{M} Y^{\mu})
\ ,
\end{equation}
where $\hat{M}$ is $2\times2$ matrix given by
\begin{equation}
\hat{M} = \left(
\begin{array}{cc}
M_{++} & 0 \\ 0 & M_{+}
\end{array} \right)
\ .
\end{equation}

The contributions to $S$ and $T$ from bileptonic gauge bosons
are given by
\begin{eqnarray}
S &=& \frac{9}{4\pi} \ln \frac{M_{++}^2}{M_+^2}
+ \frac{3}{\pi} \left[ \ln \frac{M_{++}^2}{M_{+}^2} 
- \frac{1}{18} \frac{m_Z^2}{M_+^2} \right]
\ ,
\nonumber\\
T &=& \frac{3}{16\pi m_W^2\sin^2\theta_W}
\nonumber\\
&& \times
\left[ 
  M_{++}^2 + M_{+}^2 - \frac{2M_{++}^2M_{+}^2}{M_{++}^2-M_{+}^2}
  \ln \frac{M_{++}^2}{M_{+}^2}
\right]
\nonumber\\
&& +
\frac{1}{4\pi\sin^2\theta_W}
\Bigl[
  \frac{M_{++}^2+M_{+}^2}{M_{++}^2-M_{+}^2} 
  \ln\frac{M_{++}^2}{M_+^2}
  - 2 
\nonumber\\
&& \qquad
  + 3 \tan^2\theta_W \ln\frac{M_{++}^2}{M_+^2}
\Bigr]
\ ,
\label{STbl:sf}
\end{eqnarray}
where the first terms in $S$ and $T$ are 
coming through the conventional transverse
self-energies~\cite{sasaki},
and the second terms from pinch parts~\cite{PT}.
These pinch parts are introduced to make $S$ and $T$ gauge invariant.
[The derivations of these pinch parts are given in \cite{FH}.]

Setting the $Y^{--}$ mass equal to its lower limit
and assuming that the bileptons are the only  
states additional to the SM, the singly-charged
bilepton must be heavier than
$346$ GeV to be within the 99\% CL contour of the
$S-T$ plane~\cite{Hagiwara} for Higgs mass $M_H = 300$GeV.
As another example, we put the mass of the
doubly-charged bilepton at $M(Y^{--}) = 500$GeV,
and conclude that the singly-charged partner
must lie in the mass range
$479$GeV $< M(Y^-) < 540$GeV for consistency.
If we lower $M_H$ to $100$GeV, the lower limit on
$M(Y^-)$ decreases to $324$GeV. This is an improvement
on the single-charge bilepton empirical limit ($230$GeV)
found in \cite{mudecay}.

In a grand unified model based on SU(15)
~\cite{Frampton-Lee,Frampton-Kephart} each generation of 
quarks and leptons is represented by a fundamental
{\bf 15}.  To cancel anomalies of three generations, three
generations of mirror fermions are needed.  Since these mirror
fermions obtain their masses from the VEV which breaks standard
SU(2)$_{\rm L}\times$U(1)$_{\rm Y}$ symmetry they necessarily
are close to the weak scale in mass and give significant contributions
to $S$ and $T$.
Even if we assume that members of the same SU(2)$_{\rm L}$ doublet
have degenerate masses, and hence the mirror fermions give no
contribution to $T$, they do give a very large contribution
to $S$ parameter; $S_{\rm mirror}=2/\pi$.  Then one might think that
this model is already excluded by the precision electroweak analysis?
However, there are many extra particles including bileptons and
leptoquarks in the model.  These extra particles could give
non-negligible contributions $S$ and $T$ as discussed above. 
As is easily read from Eqs.~(\ref{STbl:sf}), 
there is a negative contribution to $S$ coming from
bileptons if the singly-charged bilepton ($Y^-$) is heavier than the
doubly-charged one ($Y^{--}$).  This negative contribution can cancel
the large positive contribution coming from mirror fermions.  On the
other hand, such a mass difference of bileptons gives a large
positive contribution to the $T$ parameter.  But this could, in turn,
be canceled by a negative contribution of leptoquarks 
without affecting the $S$ parameter.

\begin{figure}
\begin{center}
\epsfxsize=2.8in
\epsfysize=2.3in
\ \epsfbox{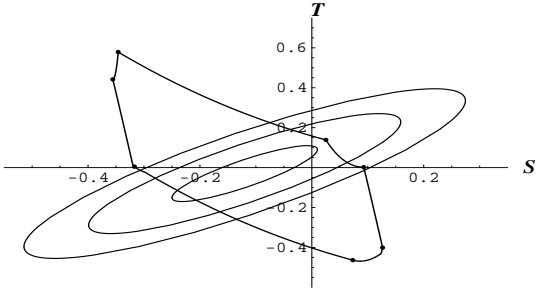}
\end{center}
\caption[]{
Illustrative region of values of $S$ and $T$ arising in SU(15)
grand unification. The mass ranges used here are: 
$425$\,GeV$\le M(Y^-)\le475$\,GeV, $250$\,GeV$\le
m_{5/3}$, $m_{-1/3}\le300$\,GeV and $300$\,GeV$\le
m_{2/3h}\le350$\,GeV. Held fixed are: $M(Y^{--})=360$\,GeV, 
$m_{2/3l}=200$\,GeV and $\alpha=\pi/4$.
}\label{fig:su15}
\end{figure}

To be specific, we show in Fig.~\ref{fig:su15} a possible region of
values of $S$ and $T$ coming from mirror fermions, bileptons and 
leptoquarks for $425$\,GeV$\le M(Y^-)\le475$\,GeV, $250$\,GeV$\le
m_{5/3}$, $m_{-1/3}\le300$\,GeV and $300$\,GeV$\le
m_{2/3h}\le350$\,GeV with $M(Y^{--})=360$\,GeV, $m_{2/3l}=200$\,GeV
and $\alpha=\pi/4$ fixed.  This demonstrates that there exists a 
region where $S$ and $T$ are acceptable: SU(15)
grand unification is not yet excluded by experiment!

The continued robustness of the standard model with respect
to more and more accurate experimental data gives tight
constraints on any attempt at ornamentation of the theory
by additional "light" physics.
The parameters $S$ and $T$ provide a very convenient
measure of compatibility with the precision electroweak data.
Here we have discussed two examples: bileptons and 
leptoquarks. 

If we identify the putative leptoquark at HERA
with a mixture of scalar doublets of SU(2)
having different hypercharge, this can also improve
agreement with data.

For bileptons we have derived a 
lower bound of $324$GeV for the singly-charged
bileptonic gauge boson, assuming that bileptons are the
only states additional to the standard model;
this improves considerably on the mass bound
available from direct measurement.     

Finally we have addressed the exaggerated reports
of the death of SU(15) grand unification due
to the large positive $S$ value from its three 
generations of mirror
fermions. Because of the simultaneous presence of 
both bileptons and 
leptoquarks in SU(15), there is an extended neighborhood
in parameter space where $S$ (and $T$)
can be acceptably small in magnitude.        

More complete details of this work are in the 
article-length paper\cite{FH}.

This work was supported in part by
the U.S. Department of Energy under
Grant No. DE-FG05-85ER-40219.

\end{document}